**Blueprinting the Future: Automatic Item Categorization using Hierarchical Zero-Shot and Few-Shot Classifiers**

Ting Wang, PhD[1]

Keith Stelter, MD[1,2]

Jenn Floyd, MS[1]

Thomas O'Neill, PhD[1]

Nathaniel Hendrix, PhD, PharmD[1]

Andrew Bazemore, MD, MPH[1]

Kevin Rode, MBA, PMP[1]

Warren P. Newton, MD, MPH[1]

1. American Board of Family Medicine, Lexington, KY
2. Mayo Clinic Health System, Mankato, MN

Conflict of Interest: None

Funding Statement: The authors received no funding to conduct this research.

Acknowledgements: None.

**Abstract**

**In testing industry, precise item categorization is pivotal to align exam questions with the designated content domains outlined in the assessment blueprint. Traditional methods either entail manual classification, which is laborious and error-prone, or utilize machine learning requiring extensive training data, often leading to model underfit or overfit issues. This study unveils a novel approach employing the zero-shot and few-shot Generative Pretrained Transformer (GPT) classifier for hierarchical item categorization, minimizing the necessity for training data, and instead, leveraging human-like language descriptions to define categories. Through a structured python dictionary, the hierarchical nature of examination blueprints is navigated seamlessly, allowing for a tiered classification of items across multiple levels. An initial simulation with artificial data demonstrates the efficacy of this method, achieving an average accuracy of 92.91% measured by the F1 score. This method was further applied to real exam items from the 2022 In-Training Examination (ITE) conducted by the American Board of Family Medicine (ABFM), reclassifying 200 items according to a newly formulated blueprint swiftly in 15 minutes, a task that traditionally could span several days among editors and physicians. This innovative approach not only drastically cuts down classification time but also ensures a consistent, principle-driven categorization, minimizing human biases and discrepancies. The ability to refine classifications by adjusting definitions adds to its robustness and sustainability.**

**Blueprinting the Future: Automatic Item Categorization using Hierarchical Zero-Shot and Few-Shot Classifiers**

Accurate item categorization is critical for mapping exam questions to the appropriate content domains delineated in the assessment blueprint. Currently, this task is achieved through either manual classification or traditional machine learning methods (Pustejovsky & Stubbs, 2012). However, manual classification is tedious, time-intensive, and prone to inconsistencies across different subject matter experts and human errors or biases. On the other hand, traditional machine learning methods requires large training data sets (Hsu, 2020). Acquiring such structured datasets demands significant resources, and the predictive accuracy is often compromised in test data due to model underfit or overfit limitations. Moreover, in the ever-evolving testing industry, blueprint categories frequently change in response to professional advancements, necessitating the provision of new training data (Pustejovsky & Stubbs, 2012).

Recently, the Generative Pretrained Transformer (GPT) based classifiers such as zero-shot and few-shot have emerged as useful tools for general text categorization (Chen & Zhang, 2021; Puri & Catanzaro, 2019). Their primary advantage is the minimization of the need for training data. They both leverages on the extensive pre-training on diverse language corpora, which equips these classifiers with a broad understanding of natural language and effectively make predictions by drawing on pre-existing linguistic patterns and relationships. Zero-shot learning with a GPT classifier involves the model making predictions or classifications without any prior specific training on the task at hand (Puri & Catanzaro, 2019). Few-shot learning, on the other hand, provides the model with a small number of examples to "prime" it for the task—this is akin to a rapid learning phase where the model quickly adapts to the specifics of the task using very limited data. (Chae & Davidson, 2023). These GPT-based approaches significantly boost the scalability of text categorization.

Nevertheless, a notable hurdle in integrating these classifiers into test development lies in navigating the hierarchical structure inherent to examination blueprints, which typically encompass multiple levels of classification. In this study, we aimed to leverage the python dictionary structure to harness the capabilities of the GPT classifiers (zero-shot and few-shot) to execute hierarchical item categorization.

*Blueprint Hierarchical Structure*

An exam blueprint, in the context of the testing industry, refers to a structured document that outlines the specific content areas that will appear on an exam. This document serves as a roadmap for test developers, ensuring that the exam aligns with the certification objectives it aims to measure. The hierarchical structure plays a pivotal role in organizing and delineating the weight and emphasis allocated to each level of categorization, thereby ensuring a well-balanced and effective assessment (Downing, 2006).

The hierarchical structure in exam blueprinting can be understood as a multi-level organization of content and skills. At the top hierarchy level, major knowledge and skill domains are identified, which are crucial competencies required for the profession. The next level delineates specific assessment objectives within these domains, followed by determining the method of assessment and establishing the amount of emphasis to allocate to each domain or objective. This hierarchical structuring ensures a systematic approach to covering all necessary material in a balanced manner, providing a clear roadmap for both test developers and test takers (Raymond, 2002; Raymond & Grande, 2019).

*Zero-Shot Classifier and Few-Shot Classifier*

The foundational mechanism of zero-shot and few-shot classifiers hinges on their ability to translate input data into a semantic space, where both the text data and category labels are represented as vectors. This transformation is facilitated by pre-trained language models, which have showcased an exceptional ability to discern complex relationships and semantic nuances within textual information. In this shared semantic space, the classifiers are equipped to make well-informed predictions, identifying the most likely category for a given input (Chen & Zhang, 2021; Srivatsa et al., 2022).

As stated above, the distinction between zero-shot and few-shot classifiers rests on their respective data requirements for training. Zero-shot classifiers operate without the need for any task-specific training data, capitalizing on the ability to infer categories based on a comprehensive, pre-existing understanding of language. This method excels when categories can be succinctly described, and the classification criteria are unambiguous. Conversely, few-shot classifiers require a minimal set of training examples to discern patterns between texts and their associated categories. This approach is particularly effective in scenarios where the text is subject to interpretation, or the classification involves subtle judgment calls. Essentially, while zero-shot learning leverages broad linguistic models for direct, clear principle-based categorization, few-shot learning fine-tunes this knowledge base with a handful of illustrative instances to navigate the complexities of nuanced text classification.

In the context of hierarchical structures, where classes might share commonalities or exhibit fine-grained differences, the few-shot learning approach is particularly advantageous. It allows for the nuanced distinction between closely related categories by leveraging a small but targeted dataset to highlight subtle differences. This method is adept at discerning the minute contrasts that define each class within the hierarchy, making it well-suited for complex classification landscapes where shared attributes might otherwise confound a more generalized model. Conversely, zero-shot learning is optimally employed for more distinct, lower-level classes where the classification criteria are explicit, and the rules are well-defined. Its strength lies in its ability to apply broad conceptual knowledge without the need for specific examples, thereby streamlining the classification process where the categorical boundaries are clear-cut (Meng et al., 2022).

By judiciously applying different classifiers to suit the nuanced demands of each hierarchical level, one can cultivate a more profound and detailed comprehension of the categorical framework. This tailored approach equips classifiers with the sophistication required to deftly navigate the intricacies of a hierarchical system, resulting in classifications that are both precise and contextually aware. Hence, the synergy between zero-shot and few-shot learning methodologies can significantly enhance the overall efficacy of hierarchical text classification.

*The Present Study*

To our knowledge, there has been no prior application of zero-shot and few-shot classifiers in categorizing questions according to a hierarchically structured blueprint. This study aims to incorporate the hierarchical organization of the blueprint into a Python-based hierarchical dictionary and utilize zero-shot and few-shot classifiers in a sequential manner across the hierarchy to achieve simultaneous classification of questions at all levels.

**Method**

The performance and application of these classifiers were examined by two studies. Study 1 illustrated the hierarchical classification framework in an artificial data example and evaluated its accuracy using the weighted accuracy metric (F1 score, ranging from 0 to 100%; closing to 100% is usually considered as "good" or "excellent", and 100% as perfect in precision and recall). Study 2 showcased an empirical study wherein an exam is mapped to a new developed blueprint. All analysis were conducted in Python 3.11. The code is available in Appendix.

*Study 1: Artificial Data Example*

In this study, we present a simulation aimed at evaluating the accuracy of the GPT classifier within a hierarchical context. Without loss of generosity, we illustrate the hierarchy framework in common words without domain knowledge definition. Suppose we have a word "Meow" needs to be classified. Our initial task is to categorize it into level 1 categories: either 'Animal' or 'Plant'. After we identified it as 'Animal', we aim to further classify it into level 2 categories: 'Mammals' or 'Birds'. Delving deeper, we then seek to distinguish it as either 'Cats' or 'Dogs' for level 3 categories. This hierarchical classification approach can be represented using Python's dictionary structure, as depicted in Figure 1. The goal is to sequentially categorize the text across three nested levels: level 1 (Animal/Plant), level 2 (Mammals/Birds within Animal; Tree/Flower within Plant) and level 3 (Dog/Cat within Mammal; Eagle/Sparrow within Birds; Oak/Pine within Tree; Rose/Tulip within Flower).

We initiated the simulation by randomly generating five words for each of the level 3 categories: rose, tulip, cat, and dog. From these 20 words sample pool, we then randomly selected 10 words. We employed the zero-shot GPT classifier in a sequential manner, utilizing it three times consecutively from higher level (level 1) to lower levels (level 2, level 3), to categorize words across three nested levels as shown above. Given that we had prior knowledge of the correct classifications, we were able to compare the results yielded by the model against the actual categories. This facilitated the calculation of classification accuracy and determination of the F1 score for each random selection. We repeated the random selection 100 times and used the final average F1 score to measure its overall accuracy.

*Study 2: Empirical Study*

In this application, we sourced 200 items from the 2022 In-Training Examination (ITE) administered by the American Board of Family Medicine (ABFM). While this exam was initially constructed based on a previous blueprint, our objective is to reclassify these items in accordance with a newly devised blueprint.

In an approach analogous to the simulation study, we began by defining the blueprint using a Python dictionary. This blueprint comprises five primary domains at level 1, and 212 specific activities categorized under level 3. Level 2 (57 categories) acts as an intermediary, encompassing various clusters of the level 3 activities. Although each domain has a conceptual definition, the question often involves a complicated application scenario that would introduce ambiguity in classification and humans must make judgment call. In this study, we utilized 10 questions for each domain as prototypical examples to train GPT-4 learn the nuanced considerations not describable in succinct languages. For level 2 and level 3 activities with distinct common definition, we utilize zero-shot classification with only a few succinct descriptive explanations. Subsequently, we undertook classification across levels 1 (few-shot classifier), 2, and 3 in a sequential manner as we did for the artificial example. After we finalized the blueprint framework setup, we address the considerations regarding text input. Specifically, we amalgamated the text from the question stem with that from the answer key to provide a comprehensive input for GPT-4. This approach ensures that GPT-4 is well-informed (keywords or test points are too short to capture the full information contained in the question stem), particularly for questions commencing with "For the following options," without being overly sidetracked by distractor options or critiques, thereby maintaining focus on the core examination point of the question.

**Results**

*Study 1: Simulation Study*

The mean F1-score, derived from 100 random selections, stands at 92.91%. Given the ubiquity of these words and categories, there was no necessity for further elucidation to the zero-shot GPT classifier. This investigation affirms that the GPT classifier (zero-shot) can be effectively deployed in a hierarchical item classification context, delivering commendable accuracy.

*Study 2: Empirical Study*

The classification process for all 200 questions, spanning three levels, was efficiently completed in a mere 15 minutes, and achieved 81% similarity in domain classification with the senior physician (KS) who provided the classification examples. Level 2 conditional similarity (given Category 1 is the same as senior physician) achieved 96.9%. Level 3 classification similarity is not comparable since the senior physician concluded that the new blueprint activity did not reflect all clinical activity shown in the question text.

**Discussion**

In this study, we leveraged the hierarchical structure provided by Python dictionaries, applying it both in a simulation to gauge accuracy and in a practical application setting. The simulation yielded an accuracy rate of 92.91%, and the real-world application enabled the classification of 200 exam questions in a swift 15 minutes, compared to days of editors and physicians' time if conducted manually.

One notable advantage of GPT classifiers lies in their remarkable convenience. In cases where categories exhibit clear-cut definitions, we can effortlessly employ a descriptive definition and a zero-shot classifier without the need for any prior training to execute the classification task. When faced with classification tasks that involve intricate considerations, a minimal set of examples for each category can be supplied, enabling the use of a few-shot classifier. This approach allows for extensive classification without the burden of lengthy training, yielding high similarity to prototype patterns within minutes. This framework has significantly enhanced efficiency when compared to traditional machine learning approaches.

In instances where a question can fall under multiple categories or none of the pre-established categories, the use of multiple label classifier proves invaluable. In these application scenarios, researchers only need to specify the maximum number of categories a question may pertain to and furnish each category with descriptive definitions. As part of our empirical study, we endeavored to classify the 200 ITE questions into six distinct clinical focus areas and six healthcare-related topics. Remarkably, the multi-label classifier accomplished this task within a mere 15 minutes, whereas the review of results by physicians spanned in days.

This methodology also showcases certain limitations, especially when criteria are applicable solely to a subset of questions within a singular activity. Addressing these distinct instances may require manual intervention, either for the manual update of classifications or to incorporate the characteristics of that specific question cluster within the context of the activity.

While manual quality control and human intervention remain imperative at the current stage, the efficiency and high similarity to physicians' classification style are invaluable in the domain of test development. Testing organizations continually grapple with the intricacies of evolving blueprints, posing considerable challenges in test formulation. Artificial Intelligence (AI) offers a convenient solution to these hurdles, diminishing human subjectivity and adhering more faithfully to established principles and definitions than can often be achieved through solely human oversight. This is particularly important as human judgment may be clouded by conflicts of interest or subtle biases not immediately discernible.

```python
hierarchy = {
    'Animal': {
        'Mammal': {
            'Dog': {},
            'Cat': {}
        },
        'Bird': {
            'Eagle': {},
            'Sparrow': {}
        }
    },
    'Plant': {
        'Tree': {
            'Oak': {},
            'Pine': {}
        },
        'Flower': {
            'Rose': {},
            'Tulip': {}
        }
    }
}
```

Figure 1. Python dictionary structure for hierarchical structure of the artificial data example.

**Appendix**

```python
def BlueprintClassifier (APIKEY = None,
                        ORG_ID = None,
                        hierarchy = None,
                        classifier = "zero-shot",
                        trainingfile = None,
                        trainingfilesheet = None,
                        classificationfile = None,
                        itemfile = None,
                        classificationfile = None,
                        outputfilename = None):
    """
    BlueprintClassifier is designed to classify text data into a hierarchical structure of labels.
    It supports two classification strategies: zero-shot and few-shot.

    Parameters:
    - APIKEY (str): Your API key for the external service.
    - ORG_ID (str): The organization ID for the external service.
    - hierarchy (dict): A nested dictionary defining the hierarchical structure of labels.
    - classifier (str): The strategy for classification: 'zero-shot' or 'few-shot'.
    - trainingfile (str): The path to the training file.
    - trainingfilesheet (str): The specific sheet in the training file.
    - classificationfile (str): The path to the classification file.
    - itemfile (str): The path to the item file.
    - outputfilename (str): The path where the output file will be saved.

    Returns:
    - DataFrame: A DataFrame containing the classified text and their corresponding hierarchical labels.

    Here's a breakdown of the function's workflow and components:

    1. Imports and Configurations:
    It imports necessary modules and sets up configurations for the SKLLM library,
    which seems to be a wrapper for machine learning models, including a ZeroShotGPTClassifier
    and FewShotGPTClassifier, likely for text classification using models like GPT-4.

    2. Reproducibility:
    Ensures the classification process is reproducible by setting a
```

```
random
    seed for NumPy, random, and PyTorch, which are libraries for 
numerical operations,
    random number generation, and machine learning, respectively.

    3. Hierarchy Processing:
    The function can extract labels from a hierarchical structure,
    which may be provided as a nested dictionary. It supports multi-
level hierarchies.

    4. Classification Functions:
    Offers two modes of classification: zero-shot and few-shot, where 
the model
    makes predictions without or with some examples, respectively. 
This is done
    using placeholder functions that should be replaced with actual 
classification code.

    5. Hierarchical Classification:
    Implements a method to classify text according to the provided 
hierarchy.
    It first predicts the top-level category and then proceeds to 
classify within
    the subcategories iteratively until it reaches the lowest level.

    6. DataFrame Construction:
    Constructs a Pandas DataFrame to store the text alongside its 
predicted categories at each level of the hierarchy.

    7. Excel Output:
    The classified results are saved to an Excel file, allowing for 
easy review and analysis.

    8. Return Value:
    Returns the DataFrame containing the classification results.
    """

    from skllm.config import SKLLMConfig
    from skllm.config import SKLLMConfig
    from skllm import ZeroShotGPTClassifier
    ## machine learning part
    from sklearn.model_selection import train_test_split
    from sklearn.linear_model import LogisticRegression
    from sklearn.metrics import accuracy_score, confusion_matrix, 
f1_score
    from sklearn.datasets import load_iris
    import numpy as np
    import pandas as pd
```

```python
from sklearn.datasets import fetch_20newsgroups
from sklearn.feature_extraction.text import TfidfVectorizer
from sklearn.naive_bayes import MultinomialNB
from sklearn.pipeline import make_pipeline
from sklearn.metrics import accuracy_score
from sklearn.linear_model import LogisticRegression
from sklearn.ensemble import RandomForestClassifier
from skllm import FewShotGPTClassifier

SKLLMConfig.set_openai_key(OPENAI_SECRET_KEY)
SKLLMConfig.set_openai_org(OPENAI_ORG_ID)

## make the classification replicable.
import random
import torch  # If you are using PyTorch

seed = 42  # or any other fixed number
np.random.seed(seed)
random.seed(seed)
torch.manual_seed(seed)

torch.backends.cudnn.deterministic = True
torch.backends.cudnn.benchmark = False

hierarchy = hierarchy

def extract_labels(hierarchy):
    level_1_labels = list(hierarchy.keys())
    level_2_labels = []
    level_3_labels = []

    for level_1_value in hierarchy.values():
        level_2_labels.extend(list(level_1_value.keys()))
        for level_2_value in level_1_value.values():
            level_3_labels.extend(list(level_2_value.keys()))

    return level_1_labels, level_2_labels, level_3_labels

def get_level2_categories(words, dictionary):
    result = []
    for word in words:
        if word in dictionary:
            result.append((list(dictionary[word])))
    return result

def get_level3_categories(dictionary, words, category1, category2):
```

```python
        """Retrieve level 3 categories for the given words based on
their level 1 and level 2 classifications."""

        level3_categories = []
        for i, word in enumerate(words):
            level1 = category1[i]
            level2 = category2[i]
            level3_category = list(dictionary[level1][level2])
            level3_categories.append(level3_category)

        return level3_categories

    def zero_shot_classify(text, category):
        # This is a placeholder function. Replace this with your zero-
shot classification code.
        gpt4_notraining = ZeroShotGPTClassifier(openai_model="gpt-4-
1106-preview")
        gpt4_notraining.fit(None, category)
        #xl = pd.ExcelFile('Training.xlsx')

        # Load a sheet into a DataFrame by its name
        #dfa = xl.parse('Sheet1')
        #gpt4_notraining =
FewShotGPTClassifier(openai_model="azure::ITE")
        #gpt4_notraining.fit(X=dfa['Text'], y=dfa['Category'])
        gpt4_notraining_pred = gpt4_notraining.predict(text)
        return gpt4_notraining_pred

    def few_shot_classify(text, category):
        # This is a placeholder function. Replace this with your zero-
shot classification code.
        #gpt4_notraining =
ZeroShotGPTClassifier(openai_model="azure::ITE")
        #gpt4_notraining.fit(None, category)
        xl = pd.ExcelFile(trainingfile)

        # Load a sheet into a DataFrame by its name
        dfa = xl.parse(sheet)
        gpt4_notraining = FewShotGPTClassifier(openai_model="gpt-4-
1106-preview")
        gpt4_notraining.fit(X=dfa['Text'], y=dfa['Category'])
        gpt4_notraining_pred = gpt4_notraining.predict(text)

        return gpt4_notraining_pred

    def classify_hierarchical_prod(text, hierarchy,
filename=filename):
```

```python
        candidates = extract_labels(hierarchy)[0]

        if classifier = "few-shot":
            category1 = few_shot_classify(text, candidates)
        elif classifier = "zero-shot":
            category1 = zero_shot_classify(text, candidates)
        else:
            print ("classifier has to be few-shot or zero-shot")

        category1simple = [text.split("Defination:")[0].strip().rstrip('.') for text in category1]

        ## obtain corresponding level 2 key
        level_2_keys = get_level2_categories(category1, hierarchy)

        ## obtain corresponding level 3 classification
        category2 = [None] * len(text)
        for i in range(len(text)):
            category2[i] = zero_shot_classify([text[i]], level_2_keys[i])[0]

        category2simple = [text.split("Defination:")[0].strip().rstrip('.') for text in category2]

        ## obtain corresponding level 3 key
        level_3_keys = get_level3_categories(hierarchy, text, category1, category2)

        ## obtain corresponding level 2 classification
        category3 = [None] * len(text)
        for i in range(len(text)):
            category3[i] = zero_shot_classify([text[i]], level_3_keys[i])[0]

        ## combine category together
        df = pd.DataFrame({
            'Text': text,
            'Category1': category1simple,
            'Category2': category2simple,
            'Category3': category3
        })

        df.to_excel(filename, index=False, engine='openpyxl')

        combined_list = list(zip(category1simple, category2simple,
```

```
category3))

        return df

    result = classify_hierarchical_prod(classificationfile['Text'], hierarchy,
                                        filename=outputfilename)

    return result
```